\documentclass[11pt]{article}
\usepackage{lineno}
\usepackage{multirow}
\usepackage{enumitem}
\usepackage{amsmath}
\usepackage{float}
\usepackage{graphicx} 
\usepackage{amssymb,amscd,amsmath,amsthm,color}
\usepackage[T1]{fontenc} 
\usepackage{subcaption}
\usepackage{enumitem}
\usepackage{mathtools}
\usepackage{tcolorbox}
\usepackage{verbatim}
\usepackage{mathrsfs}
\usepackage{physics}
\usepackage{mathtools}
\usepackage{slashed}
\usepackage{longtable}
\usepackage{jheppub}
\usepackage{hyperref}
\usepackage{cleveref}

\preprint{USTC-ICTS/PCFT-26-03}

\def\be{\begin{equation}}
\def\ee{\end{equation}}
\def\bea{\begin{eqnarray}}
\def\eea{\end{eqnarray}}
\def\dd{\mathrm d}

\def\aone{A^{(1)}}
\def\atwo{A^{(2)}}

\numberwithin{equation}{section}

\makeatletter
\g@addto@macro\bfseries{\boldmath}\g@addto@macro\bfseries{\boldmath}
\makeatother
\title{Non-supersymmetric F1-P black rings}

\author{Pavan Dharanipragada$^a$,
    Gurmeet Singh Punia$^{b,c}$,    Amitabh Virmani$^d$}
\emailAdd{pavan@physics.iitm.ac.in, gurmeet96@ustc.edu.cn, avirmani@cmi.ac.in}
\affiliation[a]{Centre for Strings, Gravitation and Cosmology,
Department of Physics,
Indian Institute of Technology Madras, Chennai, India 600036}
\affiliation[b]{
Interdisciplinary Center for Theoretical Study,
University of Science and Technology of China, Hefei, Anhui 230026, China }
\affiliation[c]{
Peng Huanwu Center for Fundamental Theory, Hefei, Anhui 230026, China}
\affiliation[d]{Chennai Mathematical Institute, 
H1, SIPCOT IT Park, Siruseri, Kelambakkam 603103 India}

\abstract{We construct singly and doubly spinning non-supersymmetric F1–P black ring solutions in five-dimensional supergravity. These black rings have regular horizons and non-zero temperature. The singly spinning configuration lies in the duality orbit of the black ring constructed by Elvang, Emparan, and Figueras, while the doubly spinning configuration is a charged extension of the black ring constructed by Chen, Hong, and Teo.  We analyze the physical properties of these solutions and the various limits they admit. In particular, the doubly spinning solution admits an extremal limit in which the entropy satisfies the relation 
$S= 2 \pi J_\phi$, thereby linking it directly to the angular momentum on the $S^2$. }
\makeatletter
\gdef\@fpheader{}
\makeatother
\begin{document}
\allowdisplaybreaks
\maketitle
\flushbottom

\section{Introduction and motivation}

Certain excited states of fundamental strings admit a semiclassical description and can be treated as macroscopic string configurations. Since they arise in perturbative string theory, their properties can be analyzed in detail, while their extended nature allows them to source spacetime fields and generate corresponding supergravity backgrounds. These  complementary descriptions developed in 
\cite{Dabholkar:1989jt, Dabholkar:1990yf, Dabholkar:1995nc, Callan:1995hn,
Sen:1995in} have been central to progress in black-hole microphysics \cite{Lunin:2001fv, Dabholkar:2004yr}.  The discovery of five-dimensional black rings \cite{Emparan:2001wn, Emparan:2004wy, Elvang:2004rt,  Emparan:2006mm, Emparan:2008eg}, especially the dipole black ring by Emparan \cite{Emparan:2004wy},  reinvigorated these ideas further. By adding charges on the dipole black ring,  it became possible to construct supersymmetric two-charge black rings \cite{Elvang:2004xi}\footnote{In the earlier developments charged black rings were often viewed as supertubes dimensionally reduced along the tube direction \cite{Elvang:2003mj, Elvang:2004xi}.} (often called small black rings)  and their finite temperature cousins. Soon afterwards, it was understood that   the Bekenstein-Hawking-Wald entropy of the small black ring agrees with the count of certain supersymmetric states of fundamental string up to an overall normalization \cite{Dabholkar:2005qs, Dabholkar:2006za, Sen:2009bm}.

The subject has seen renewed attention in the past few years. As in other studies of the microscopic structure of black holes in string theory, in references \cite{Dabholkar:2005qs, Dabholkar:2006za, Sen:2009bm}, an index was computed in a  weakly coupled string theory. How should the same index be computed on the gravity side? For a long time we did not know how to answer this question. Much of the recent interest stems from proposals to evaluate supersymmetric indices directly on the gravity side \cite{Cabo-Bizet:2018ehj, Iliesiu:2021are}. The essential idea is to work at finite temperature and introduce a chemical potential for the angular momentum, which effectively inserts a factor of $(-1)^F$ in the trace computed by the gravitational path integral.
This new viewpoint on the BPS entropy of supersymmetric black holes is non-trivial even at the classical level. Notably, the match between the two sides is not between the entropies themselves, but between the entropy of an extremal black hole and that of a non-extremal black hole supplemented by a term proportional to the angular momentum it carries \cite{Iliesiu:2021are}.

Applying these ideas to black holes in five dimensions is a subject of much discussion. At least three different proposals have been put forward for identifying the index saddles for black rings \cite{Bandyopadhyay:2025jbc, Cassani:2025iix, Boruch:2025sie}. The three proposals agree on several aspects but also differ on several other aspects. A key reason for the differences is  in  the way they treat the two five-dimensional angular momenta. In \cite{Bandyopadhyay:2025jbc}, the small black ring for which the index saddle was constructed carries only two electric charges  $Q_1, Q_2$, and only one angular momentum $J_\psi$ along the $S^1$ of the ring. The  identified saddle has $Q_1, Q_2, J_\psi$ and in  addition, it has purely imaginary angular momentum $J_\phi$ on the $S^2$ cross-section of the ring. 
The index saddle is constructed by analytically continuing a non-extremal solution, \emph{with only the 
 $J_\phi$
  angular momentum continued to purely imaginary values.} The construction has several parallels  to the construction of the index saddles for small black holes \cite{Chowdhury:2024ngg, Chen:2024gmc}.

The 4d-5d connection \cite{Gaiotto:2005xt, Gaiotto:2005gf} (specifically, how the supersymmetric black ring solutions are written in terms of harmonic functions \cite{Gauntlett:2004wh, Bena:2007kg}), then implies that the total momentum charge as captured by the $H_0$ harmonic function\footnote{In a standard $N=2$ supergravity notation this harmonic function is denoted $H_0$. In the Bena-Warner notation this function is denoted $M$.} is complex.  This differs from the analysis of \cite{Boruch:2025sie}. The approach of \cite{Boruch:2025sie} is strongly anchored on a four-dimensional analysis \cite{Boruch:2025qdq, Boruch:2025biv}, where all total charges as captured by  the harmonic functions are taken to be real. Treatment in~\cite{Cassani:2025iix}  is similar to~\cite{Bandyopadhyay:2025jbc}, however, a detailed comparison is yet to be done.\footnote{As already pointed out in \cite{Cassani:2025iix} a detailed comparison is not so straightforward. The solutions of \cite{Bandyopadhyay:2025jbc}
have running scalars and there is no limit in which they reduce to the solutions of minimal supergravity
considered in \cite{Cassani:2025iix}. However, in principle, a comparison is possible as the analysis of  \cite{Cassani:2025iix} can be generalised to 5d theories with vector multiplets.}

As the subject develops, it is important to do more examples to compare and contrast different analyses. A key step in the analysis of \cite{Bandyopadhyay:2025jbc} is the construction of a two-charge non-extremal black ring solution  with two independent rotations. To extend the analysis of \cite{Bandyopadhyay:2025jbc} to precisely the set-up where the index computations of \cite{Dabholkar:2005qs, Dabholkar:2006za, Sen:2009bm} apply, we need to construct a charged version of the doubly spinning dipole black ring. The key aim of this paper is to present precisely such a solution: a smooth, Lorentzian, non-extremal, two-charge, doubly spinning dipole black ring. Since the doubly spinning dipole black ring is a fairly cumbersome solution, construction  of the requisite charged solution is a task in itself. The final solution has several parameters.   In a separate paper \cite{Dharanipragada:2026dji}, we analyse the analytic continuation  that gives the index saddle for the F1-P black ring.

The rest of the paper is organized as follows. In section \ref{sec:dualities}, we discuss dualities that add F1-P charges to dipole black rings. We present the final answer as a recipe that can be applied to either singly spinning or doubly spinning dipole black ring solution. In section \ref{sec:EEF}, we apply the dualities to Emparan's dipole black ring \cite{Emparan:2004wy} (with a single dipole charge) to generate the singly spinning non-supersymmetric F1-P black ring. This solution is in the duality orbit of a previously constructed black ring by Elvang, Emparan, and Figueras \cite{Elvang:2004xi}.  In section \ref{sec:Teo}, we apply the same dualities to the 
doubly spinning dipole black ring of Chen, Hong, and Teo \cite{Chen:2012kd}. We discuss physical properties of the charged solution, including various limits it  admits. In particular, we highlight that the doubly spinning solution admits an extremal limit where the entropy is related to its $S^2$ angular momentum by $S=2 \pi J_\phi$. This feature is analogous  to the rotating black holes whose analytic continuation gives index saddles for small black holes \cite{Chowdhury:2024ngg, Chen:2024gmc}. We close in section \ref{sec:conclusions} with a brief discussion.   
 


\section{Dualities to add F1-P charges}
\label{sec:dualities}

 Dipole black rings \cite{Emparan:2004wy, Chen:2012kd} are solutions of a five-dimensional theory with Lagrangian
\be\label{e2daction}
{\cal L} =
R-\frac{1}{2}(\nabla \widetilde \phi)^2-\frac{1}{12}e^{-\frac{2 \sqrt{2}}{\sqrt{3}} \widetilde \phi} H_{\mu \nu \rho} H^{\mu \nu \rho}\, ,
\ee
where $H= dB$.  Lagrangian \eqref{e2daction} can be interpreted as the NS–NS sector of low-energy string theory and admits the fundamental string as a solution. Our aim is to add two charges to dipole black ring solutions so that the resulting configurations admit an interpretation as F1–P black rings in string theory. We achieve this by embedding the five-dimensional theory into a six-dimensional theory and applying a sequence of duality transformations.  The uplift step is somewhat non-trivial due to the presence of the B-field and the scalar $\widetilde \phi$ in the Lagrangian \eqref{e2daction}.  We proceed as follows.

A suitable truncation 
of the low energy NS-NS sector of superstring theory compactified on $T^4$ yields  a six-dimensional theory containing a  metric, an antisymmetric two-form field $B_{MN}$, and a dilaton $\Phi$.  In the string frame, the action takes the form, 
 \be 
 S_{6S} = \frac{1}{16\pi G_6} \int \dd^6x \sqrt{-G^{(S)}} e^{-2\Phi}\left[ R^{(S)} + 4(\nabla \Phi)^2  - \frac{1}{12}  H_{MNP}H^{MNP}\right], \label{action_6sd}
 \ee
 where $H = dB$.  The Einstein frame metric is obtained via
\be \label{eq:string-Einstein}
 G_{MN}^{(E)} = e^{-\Phi} G^{(S)}_{MN},
\ee 
and the Einstein frame action reads
\be
 S_{6E} = \frac{1}{16\pi G_6} \int \dd^6x \sqrt{-G^{(E)}} \left[ R^{(E)} -(\nabla \Phi)^2  - \frac{1}{12}  e^{-2\Phi}H_{MNP}H^{MNP}\right]. \label{action_6ed}
 \ee

 We are interested in charged solutions of the theory obtained by a further dimensional reduction  of the action \eqref{action_6ed} on an $S^1$.  Using the following ansatz for the Kaluza-Klein reduction of the metric,
 \be \label{KK6d}
\dd s^2_{6E} = e^{\frac{1}{\sqrt{6}}\chi} \dd s_{5E}^2 + e^{-\frac{\sqrt{3}}{\sqrt{2}} \chi} (\dd z + A^{(1)})^2,
\ee
we obtain the five-dimensional Einstein frame theory. The 
NS-NS two-form $B_{M N}(x,z)$ is reduced as,
\be
B(x,z) = B(x) + A^{(2)}(x) \wedge \dd z,
\ee 
where $B(x)$ is a two-form in five dimensions and $A^{(2)}(x)$ is a one-form.  The resulting five-dimensional Einstein frame action is
\be
 S = \frac{1}{16\pi G_5} \int \dd^5 x \sqrt{-g}~{\cal L}, \label{action_5d} \\
\ee
\be \label{5D_action}
{\cal L} = R - \frac{1}{2}(\nabla \chi)^2   - (\nabla \Phi)^2 - \frac{1}{12}e^{-\frac{\sqrt{2}}{\sqrt{3}}\chi - 2\Phi} H^2 -\frac{1}{4}e^{-\frac{2 \sqrt{2}}{\sqrt{3}}\chi}\left(F^{(1)}\right)^2
-\frac{1}{4}e^{\frac{\sqrt{2}}{\sqrt{3}}\chi -2\Phi}\left(F^{(2)}\right)^2,
 \ee
with the field strengths defined by 
\be
H = \dd B - \dd A^{(2)} \wedge A^{(1)},
\ee
 $F^{(1)}= \dd A^{(1)}$  and $F^{(2)} = \dd A^{(2)}$.

How is the Lagrangian \eqref{e2daction} embedded in \eqref{5D_action}? There are several possible embeddings. We are interested in the embedding where the B-field in \eqref{e2daction} is identified with the B-field in \eqref{5D_action}. This requires an appropriate identification of the scalar fields.

We begin by matching the coefficients in the exponential factors multiplying $H^2$. Specifically, we require
\be
-\frac{\sqrt{2}}{\sqrt{3}}\chi - 2\Phi = -\frac{2 \sqrt{2}}{\sqrt{3}} \widetilde \phi.
\ee
This implies, 
\be
\widetilde \phi = \frac{\sqrt{3}}{\sqrt{2}} \Phi + \frac{1}{2}\chi.
\ee
Next, we introduce a second scalar field. We choose 
$\widetilde \psi$ such that the kinetic terms for the scalar fields take the same canonical form as in \eqref{5D_action}:
$- \frac{1}{2}(\nabla \widetilde \phi)^2   - (\nabla \widetilde \psi)^2$. This is  achieved by defining 
\be
\widetilde \psi = \frac{1}{2} \Phi - \frac{\sqrt{3}}{2\sqrt{2}}\chi.
\ee

With these field redefinitions, the Lagrangian \eqref{5D_action} can be written as
\be
{\cal L} = R - \frac{1}{2}(\nabla \widetilde \phi)^2   - (\nabla \widetilde \psi)^2 - \frac{1}{12}e^{-\frac{2 \sqrt{2}}{\sqrt{3}} \widetilde \phi} H^2 -\frac{1}{4}e^{-\frac{\sqrt{2}}{\sqrt{3}}\widetilde \phi + 2 \widetilde \psi}\left(F^{(1)}\right)^2
-\frac{1}{4}e^{-\frac{\sqrt{2}}{\sqrt{3}}\widetilde \phi - 2 \widetilde \psi}\left(F^{(2)}\right)^2.
 \ee
The truncation to the simplified Lagrangian \eqref{e2daction} is then obtained by setting
\be \label{A1-A2-0}
A^{(1)} = A^{(2)} = 0,  \qquad \widetilde \psi = 0.
\ee

In this truncation, the fields $\Phi$  and $\chi$  are related  via 
\be \label{chi-Phi}
\Phi = \frac{\sqrt{3}}{\sqrt{2}}\chi ,
\ee
which in turn implies  
\be
\widetilde \phi = \frac{2\sqrt{2}}{\sqrt{3}} \Phi.
\ee
The field $\widetilde \phi$ in \eqref{e2daction} is therefore directly related to the string theory dilaton.

The six-dimensional Einstein frame metric \eqref{KK6d}, in the truncation defined by \eqref{A1-A2-0}, is therefore
 \be
\dd s^2_{6E} = e^{\frac{1}{3}\Phi} \dd s_{5E}^2 + e^{-\Phi} \dd z^2,
\ee
and the corresponding string frame metric is
\be \label{metric_6D_seed}
ds^2_{6S} = e^{\frac{4}{3}\Phi} ds_{5E}^2 + dz^2.
\ee

Action \eqref{action_6sd} is identical to the low-energy NS-NS sector of string theory. T-duality is a symmetry of this action \cite{Hassan:1991mq, Maharana:1992my}. We can therefore use T-duality to add charges under $\aone$ and $\atwo$ to solutions of 
the theory \eqref{e2daction}. We do so as follows:
\begin{enumerate}
\item We start with a solution to the truncated Lagrangian \eqref{e2daction} and uplift it to six-dimensions via \eqref{metric_6D_seed}.  A Lorentz boost with boost parameter $\delta_2$ along the $z$-direction,  
\bea
t&=&  t' \cosh \delta_2 +  z' \sinh\delta_2,   \label{boost-1-1}\\ 
z&=&  z'  \cosh \delta_2 +  t' \sinh\delta_2. \label{boost-1-2}
\eea
gives a solution with linear momentum in the compact direction. 

\item 
We apply T-duality along the $z'$ direction. The  rules  are (for ease of notation we call $z'=s$):
 \begin{align}
 G'_{ss}&=\frac{1}{G_{ss}}, &
 e^{2\Phi'}&=\frac{e^{2\Phi}}{G_{ss}}, \\ 
 G'_{\mu s}&=\frac{B_{\mu
 s}}{G_{ss}},& B'_{\mu s}&=\frac{G_{\mu s}}{G_{ss}},\\
G'_{\mu \nu}&=G_{\mu \nu}-\frac{G_{\mu s}G_{\nu s}-B_{\mu s}B_{\nu  
s}}{G_{ss}}, &
B'_{\mu \nu}&=B_{\mu \nu}-\frac{B_{\mu s}G_{\nu s}-G_{\mu s}B_{\nu  
s}}{G_{ss}}.
 \end{align}
This step converts the momentum charge into F1 charge. 
\item  
Finally, we perform another boost in the   $z'$ direction with boost parameter $\delta_1$,
\bea
t' &=& t'' \cosh \delta_1  + z''  \sinh\delta_1, \\ 
 z'&=&  z''  \cosh \delta_1  + t''  \sinh\delta_1.
\eea
This boost gives a solution with linear momentum in the compact direction. 
\end{enumerate}
The resulting configuration is a solution to the equations of motion obtained from \eqref{action_6sd}. Dimensional reduction to the five-dimensional Einstein frame gives the solution of interest. The solution carries an F1 charge (related to parameter $\delta_2$) under $\atwo$ and a momentum charge (related to parameter $\delta_1$) under $\aone$.

Starting with  a general configuration, where the 5d Einstein frame metric is the form,
\be \label{sol-1}
\dd s^2 =  g_{tt} (\dd t + \omega_{\psi} \dd \psi + \omega_{\phi}  \dd \phi)^2 + 
g_{\psi \psi} (\dd \psi + \omega_{\psi \phi} \dd \phi)^2 + g_{\phi \phi} \dd \phi^2  + g_{xx} \dd x^2 + g_{yy} \dd y^2, 
\ee
and the B-field is of the form,
\be \label{sol-2}
B = B_{t \phi} \dd t \wedge \dd \phi + B_{t \psi} \dd t \wedge \dd \psi + B_{\phi \psi} \dd \phi \wedge \dd  \psi,
\ee
together with the dilaton $\Phi$, the transformed configuration  in the 5d Einstein frame is
\bea
 \dd \check s^2   &= & (h_1 h_2)^{-2/3} g_{tt} (\dd t + \check \omega_{\psi} \dd \psi  + \check \omega_{\phi}  \dd \phi)^2  \nonumber \\  & & + ~ (h_1 h_2)^{1/3} \left( g_{\psi \psi} (\dd \psi + \omega_{\psi \phi} \dd \phi)^2 + g_{\phi \phi} \dd \phi^2 + g_{xx} \dd x^2 + g_{yy} \dd y^2 \right), \\
\check \omega_{\psi} &=& c_1 c_2 \omega_{\psi} + s_1 s_2  B_{t \psi}, \\
\check \omega_{\phi} &=&  c_1 c_2 \omega_{\phi} + s_1 s_2  B_{t \phi},
\eea
where
\begin{align}
h_1 &=  c_1^2 + s_1^2 e^{4\Phi/3} g_{tt}, \\
h_2 &=  c_2^2 + s_2^2 e^{4\Phi/3} g_{tt}.
\end{align}
The transformed B-field is
\bea
\check   B_{t \phi} &=& \frac{1}{h_2} \left( c_1 c_2 B_{t \phi} - s_1 s_2 \omega_\phi  e^{4\Phi/3} g_{tt}\right)  ,\\
\check   B_{t \psi} &=&  \frac{1}{h_2} \left( c_1 c_2 B_{t \psi} - s_1 s_2 \omega_\psi  e^{4\Phi/3} g_{tt} \right),\\
\check   B_{\phi \psi} &=& \frac{1}{h_2} \left( c_2^2  B_{\phi \psi} + s_2^2 (B_{\phi \psi} - B_{t \psi} \omega_\phi + B_{t \phi} \omega_\psi )e^{4\Phi/3} g_{tt} \right).
\eea
The scalars are
\begin{align}
e^{2 \check \Phi} &=  h_2^{-1} e^{2 \Phi}, \\
e^{- \sqrt{\frac{3}{2}}\check \chi} &=  \frac{h_1}{\sqrt{h_2}} e^{-\Phi},
\end{align}
and finally the two vectors are
\begin{align}
\check A {}^{(1)}_t & =   h_1^{-1} \left(1+ e^{4\Phi/3} g_{tt} \right) c_1 s_1,  \\
\check A {}^{(2)}_t &=  h_2^{-1} \left(1+ e^{4\Phi/3} g_{tt} \right) c_2 s_2,\\
\check A {}^{(1)}_\phi & =  - h_1^{-1} \left( c_1 s_2  B_{t\phi} - s_1 c_2 \omega_\phi  e^{4\Phi/3} g_{tt} \right), \\
\check A {}^{(2)}_\phi &=  - h_2^{-1} \left( c_2 s_1 B_{t\phi}  - s_2 c_1 \omega_\phi  e^{4\Phi/3} g_{tt} \right),\\
\check A {}^{(1)}_\psi & =  - h_1^{-1} \left( c_1 s_2  B_{t\psi} - s_1 c_2 \omega_\psi  e^{4\Phi/3} g_{tt} \right), \\
\check A {}^{(2)}_\psi &=  - h_2^{-1}  \left( c_2 s_1 B_{t\psi}  - s_2 c_1 \omega_\psi  e^{4\Phi/3} g_{tt} \right).
\end{align}
These expressions provide a general recipe for adding F1 and P charges to any solution of the theory \eqref{e2daction}.

\section{Singly spinning non-supersymmetric F1-P black ring}
\label{sec:EEF}

In this section we apply the dualities to Emparan's dipole black ring \cite{Emparan:2004wy} to generate the singly spinning non-supersymmetric F1-P black ring.  

\subsection{The solution and physical properties}

For Emparan's dipole black ring we use slightly different notation compared to the original paper. This is to facilitate comparison with the doubly spinning solution we will discuss in section \ref{sec:Teo}. Our presentation is closely related to that of \cite{Chen:2012kd}. The metric takes the form, 
\bea
\dd s^2&=&-\frac{F(y)}{F(x)}\,\bigg[\frac{ H(x)}{ H(y)}\bigg]^{\frac{1}{3}}\left(\dd t+\omega_\psi \dd\psi\right)^2+\frac{2\varkappa^2}{(x-y)^2}F(x)\left[ H(x) H(y)^2\right]^{\frac{1}{3}}\,\nonumber \\[1mm]
&&\times\left\{-\frac{G(y)\,\dd\psi^2}{F(y) H(y)}+\frac{G(x)\,\dd\phi^2}{F(x) H(x)}+\frac{1}{1-a^2}\bigg[\frac{\dd x^2}{G(x)}-\frac{\dd y^2}{G(y)}\bigg]\right\},
\eea
with
\be
\omega_\psi=-\sqrt{\frac{2a(1+a)(a+c)}{1-a}}\,\frac{\varkappa(1+c)(1+y)}{F(y)},
\ee
where we have normalised the angular coordinates so that they have canonical periodicity. We have also used the `balance condition' of \cite{Emparan:2004wy} implicitly in writing the solution. The radial coordinate $y$ takes the range $- \infty < y \le -1$, and the polar coordinate $x$ on the $S^2$ takes the range $- 1 \le x \le 1$.  The functions $F, G$ and $H$ are given as,
\bea
F(x)&=&1+ac+(a+c)x, \\ G(x) &=& (1-x^2) (1+cx),  \\ H(x)&=&1-ac-(a-c)x.
\eea
The solution depends on three independent parameters $a, c, \varkappa$, subject to the constraints,
\begin{align}
& 0 \le c \le a < 1,   & & \varkappa > 0.
\end{align}
Roughly speaking, $\varkappa$ sets the scale of the solution. The difference between the parameters $a$ and $c$,  $a-c$, is related to the dipole charge and the parameter $a$ is related to the $S^1$ rotation of the ring. 
The B-field supporting the solution is, 
\be
B_{t\psi} = - \sqrt{\frac{2a(1-a)(a-c)}{1+a}}\,\frac{\varkappa(1+c)(1+y)}{H(y)}, 
\ee
and the dilaton is,
\be
e^{\widetilde \phi} = e^{\frac{2\sqrt{2}}{\sqrt{3}} \Phi} = \left[\frac{H(x)}{H(y)}\right]^\frac{\sqrt{2}}{\sqrt{3}}.
\ee

The parameters $\mu, \nu, R$ of \cite{Emparan:2004wy}\footnote{The balance condition is solved, so there is no $\lambda$ in our presentation.} are related to parameters $a, c,  \varkappa $ as,
\bea \label{dictionary-EEF-1}
a=\frac{\mu+\nu}{1+\mu\nu}\,,\qquad c=\nu\,,\qquad \varkappa=R\,\sqrt{\frac{1-\mu^2}{2(1+\nu^2+2\mu\nu)}}.
\eea
The coordinates $(\psi, \phi)$ are related to $(\psi_E,\phi_E)$  used in \cite{Emparan:2004wy} as, 
\be \label{dictionary-EEF-2}
(\psi,\phi)=\sqrt{\frac{1+\nu^2+2\mu\nu}{1-\mu^2}}\,(-\psi_E,\phi_E).
\ee

We can now write the two-charge solution. We first observe that
\be
e^{4\Phi/3} g_{tt} = - \left[\frac{H(x)}{H(y)}\right]^\frac{2}{3} \cdot \frac{F(y)}{F(x)}\,\bigg[\frac{ H(x)}{ H(y)}\bigg]^{\frac{1}{3}} = - \frac{H(x) F(y) }{H(y) F(x)}. 
\ee
As  a result, 
\be
h_i = c_i^2 + s_i^2 e^{4\Phi/3} g_{tt} 
= c_i^2 - s_i^2 \frac{H(x) F(y) }{H(y) F(x)} 
= 1 + \frac{2a(1-c^2)(x-y)s_i^2}{H(y) F(x)}. 
\ee
The metric for the two-charge singly spinning dipole black ring takes the form,
\bea  \label{metric-EEF}
 \dd \check s^2   &= &- (h_1 h_2)^{-2/3} \frac{F(y)}{F(x)}\,\bigg[\frac{ H(x)}{ H(y)}\bigg]^{\frac{1}{3}} (\dd t + (c_1 c_2 \omega_{\psi} + s_1 s_2  B_{t \psi}) \dd \psi)^2  \nonumber \\[1mm]  & & + ~ (h_1 h_2)^{1/3} \frac{2\varkappa^2}{(x-y)^2}F(x)\left[ H(x) H(y)^2\right]^{\frac{1}{3}}\,\nonumber \\[1mm]
&&\times\left\{-\frac{G(y)\,\dd\psi^2}{F(y) H(y)}+\frac{G(x)\,\dd\phi^2}{F(x) H(x)}+\frac{1}{1-a^2}\bigg[\frac{\dd x^2}{G(x)}-\frac{\dd y^2}{G(y)}\bigg]\right\}.
\eea
The transformed B-field is,
\be
\check B_{t \psi} =  \frac{1}{h_2} \left( c_1 c_2 B_{t \psi} + s_1 s_2 \omega_\psi  \frac{H(x) F(y) }{H(y) F(x)} \right).
\ee
The scalars are,
\begin{align}
e^{2 \check \Phi} &= \frac{1}{h_2} \frac{H(x)}{H(y)}, \\
e^{- \sqrt{\frac{3}{2}}\check \chi} &=  \frac{h_1}{\sqrt{h_2}} \left[\frac{H(y)}{H(x)}\right]^\frac{1}{2},
\end{align}
and finally the two vectors are, 
\begin{align}  \label{vec1-EEF}
\check A {}^{(1)}_t & =   h_1^{-1} \left(1- \frac{H(x) F(y) }{H(y) F(x)} \right) c_1 s_1,  \\
\check A {}^{(2)}_t &=  h_2^{-1} \left(1- \frac{H(x) F(y) }{H(y) F(x)}\right) c_2 s_2,\\
\check A {}^{(1)}_\psi & = - h_1^{-1} \left(c_1 s_2  B_{t\psi} + s_1 c_2 \omega_\psi  \frac{H(x) F(y) }{H(y) F(x)} \right), \\
\check A {}^{(2)}_\psi &=  - h_2^{-1}  \left(c_2 s_1 B_{t\psi} + s_2 c_1 \omega_\psi  \frac{H(x) F(y) }{H(y) F(x)}\right). \label{vec2-EEF}
\end{align}

The two-charge solution \eqref{metric-EEF}--\eqref{vec2-EEF} is in the duality orbit of a previously constructed black ring by Elvang, Emparan, and Figueras \cite{Elvang:2004xi}. This can be seen as follows: first we can dualise the B-field to a one-form $A^{(3)}$. The solution can then we interpreted as a solution to five-dimensional U(1)$^3$ theory. Any solution of U(1)$^3$ theory can be uplifted to IIB theory (see, e.g.,  equations (50) and (51) of \cite{Emparan:2006mm}). We can do the IIB uplift using $A^{(3)}$ as the Kaluza-Klein vector field to six-dimensions. The resulting configuration is exactly the same as the one given in \cite[section 5.2]{Elvang:2004xi}.

The asymptotically flat nature of the solution  \eqref{metric-EEF}--\eqref{vec2-EEF} can be made manifest by changing coordinates
\bea
x&=& -1 + \frac{4 \varkappa^2}{r^2} (1-c) \cos^2 \theta, \\
y &=& -1 - \frac{4 \varkappa^2}{r^2} (1-c) \sin^2 \theta.
\eea
The ADM mass of the black ring \eqref{metric-EEF}--\eqref{vec2-EEF} is 
\be \label{mass-EEF}
 M = \frac{\pi  \varkappa ^2}{\left(1-a^2\right) G_5}  \left\{(a + c) (1+a) + a(1 + c) (\cosh 2 \delta_1 + \cosh 2\delta_2-1)\right\},
\ee
    and the $S^1$ angular momentum $J_\psi$ is
\be
  J_\psi = \frac{\pi  \varkappa ^3 (1+c)}{G_5 (1-a^2)} \left[ c_1 c_2 (1+a)^2 \sqrt{\frac{2 a (a+c)}{1-a^2}} + s_1 s_2 (1-a)^2 \sqrt{\frac{2 a (a-c)}{1-a^2}} \right].
  \ee
The $S^2$ angular momentum $J_\phi$ is zero. The P- and F1- charges are, respectively,
\bea \label{charge-1}
\mathbf{Q}_{1} &=& \frac{1}{16 \pi G_5} \int_{S^3_\infty} e^{-\frac{2 \sqrt{2}}{\sqrt{3}}\chi} \star_5 F^{(1)}=  
\frac{2 \pi \varkappa^2 a(1+c)}{G_5(1-a^2)} s_1 c_1, \\ [2mm]
\mathbf{Q}_{2} &=&\frac{1}{16 \pi G_5} \int_{S^3_\infty} e^{\frac{\sqrt{2}}{\sqrt{3}}\chi -2\Phi} \star_5 F^{(2)} = \frac{2 \pi \varkappa^2 a(1+c)}{G_5(1-a^2)}s_2 c_2, \label{charge-2}
\eea
and the dipole charge is\footnote{We use epsilon convention 
$\epsilon_{t y x \psi \phi} > 0$
and use Mathematica package \texttt{diffgeo.m} function \texttt{HodgeStar} to perform Hodge dualities. This means,
\be
(\star_5 H)_{\alpha \beta} = H_{\mu \nu \rho} \epsilon^{\mu \nu \rho}{}_{\alpha \beta}.
\ee
As far as this specific Hodge duality is concerned, we could equally well use the function \texttt{HodgeStarPolchinski},
\be
(\star_5 H)_{\alpha \beta} = \epsilon_{\alpha \beta}{}^{\mu \nu \rho} H_{\mu \nu \rho}.
\ee
For the sphere integral we use $d x \wedge d\phi$ as the orientation.}
\be \label{dipole-charge-EEF}
q =\frac{1}{2 \pi} \int_{S^2} e^{-\frac{\sqrt{2}}{\sqrt{3}}\chi - 2\Phi} \star_5 H =  \frac{2\varkappa}{\sqrt{1-a^2}}\left[\sqrt{2 a (a-c)} \, c_1 c_2 +  \sqrt{2 a (a+c)} \, s_1 s_2 \right].
\ee
Note that even when the seed solution has zero dipole charge, i.e., $a=c$, the charged solution has a non-zero dipole charge.

The horizon is at $y=-1/c$. The horizon area is,
\be \label{area-EEF}
\mathcal{A}_H = 16 \pi ^2 \varkappa ^3  c \left[ \frac{\sqrt{2a (a+c)}}{1-a} c_1 c_2 -  \frac{\sqrt{2a (a-c)}}{1+a} s_1 s_2\right],
\ee
 the temperature is,
\be
T_H = \frac{1}{4 \pi  \varkappa }  \left[ \frac{\sqrt{2a (a+c)}}{1-a} c_1 c_2 -  \frac{\sqrt{2a (a-c)}}{1+a} s_1 s_2 \right]^{-1},
\ee
and the $\Omega_\psi$ angular velocity of the horizon is,
\be
 \Omega_\psi = \frac{1}{\varkappa}  \left[\sqrt{\frac{2(1+a) (a+c)}{ a(1-a)}}c_1 c_2  -  \sqrt{\frac{2(1-a) (a-c)}{a (1+a)}} s_1 s_2 \right]^{-1}.
 \ee

It is instructive to compare expressions \eqref{mass-EEF}--\eqref{area-EEF} to the expressions given in \cite[section 5.2.2]{Elvang:2004xi}. Using the dictionary \eqref{dictionary-EEF-1}--\eqref{dictionary-EEF-2} it can be readily checked that all expressions match. Though, of course, the interpretations for the electric charges and the dipole charge in terms of the underlying branes are different.

\subsection{Supersymmetric limit to a small black ring}
\label{sec:EEF-BPS}
The BPS limit requires taking the boost parameters to infinity $\delta_1,\delta_2 \to \infty$, while keeping the charges fixed. This limit was discussed in \cite{Elvang:2004xi}, where it was argued that we need to take $\mu, \nu \to 0$ keeping $\mu/\nu$ fixed as we take $\delta_1,\delta_2 \to \infty$. From the dictionary \eqref{dictionary-EEF-1}, we see that we need to take $a$ and $c$ to zero, keeping the ratio $c/a$ fixed.  One convenient way to implement this is to set,
\be \label{introducing-alpha}
c = a \alpha \qquad \mbox{and}  \qquad \delta_i=\frac12\sinh^{-1}\bigg(\frac{Q_i}{4\varkappa^2 a}\bigg),
\ee
and take $a \to 0$. 
In this limit, 
\be
M= \frac{\pi}{4 G_5} \left( Q_{1} + Q_{2} \right),
\ee
and 
\be
\mathbf{Q}_{1,2} = \frac{\pi}{4 G_5}  Q_{1,2}.
\ee

The solution is now parameterized by $Q_1, Q_2$,  $\alpha$, and  $\varkappa$. In the BPS limit, the dipole charge is expressed in terms of the parameter $\alpha$ as
\begin{equation} \label{dipole-alpha}
    q=\frac{1}{2\sqrt{2} \varkappa} \sqrt{Q_1 Q_2} (\sqrt{1+\alpha}+ \sqrt{1-\alpha}),
\end{equation}
with $0 \le \alpha \le 1$. The signs of $Q_1, Q_2, q$ are positive in our conventions.   The function $(\sqrt{1+\alpha}+ \sqrt{1-\alpha})$ ranges between $[\sqrt{2},2]$. Therefore, 
\be \label{bound}
\frac{1}{2 \varkappa} \sqrt{Q_1 Q_2} \le q \le \frac{1}{\sqrt{2} \varkappa} \sqrt{Q_1 Q_2}.
\ee
The dipole charge is bounded from both above and below.

In the BPS limit, the metric functions $F,G,H$ become 
\begin{align}
F(\xi) & \to 1,& H(\xi)& \to 1, & G(\xi) & \to 1-\xi^2.
\end{align}
As a result, the metric becomes,
\bea
	\dd s_5^2&=&-(h_1 h_2)^{-2/3}(\dd t+\omega)^2+(h_1 h_2)^{1/3}\dd s_4^2,\\
	\label{bpsmetric}
	h_i&=&1+ \frac{Q_i}{4 \varkappa^2}(x-y), \quad 
	\omega = - \frac{q}{2}(1+y) \dd \psi.
\eea
The four-dimensional base metric $\dd s_4^2$ is flat space in ring coordinates,
\begin{equation}
	\dd s_4^2= \frac{2\varkappa^2}{(x-y)^2} \left[\frac{\dd y^2}{y^2-1} + (y^2-1)\dd\psi^2 + \frac{\dd x^2}{1-x^2} + (1-x^2) \dd\phi^2 \right].
	\label{ds4}
\end{equation}
The remaining fields supporting the solution are,
\begin{align}
	& B =- \frac{1}{2 h_2 } q(1+y) \dd t \wedge \dd \psi,& &
e^{2\Phi} =\frac1{h_2},& &
    e^{- \sqrt{\frac{3}{2}}\chi} =  \frac{h_1}{\sqrt{h_2}}, \\
	& A^{(1)} =\dd t - h_1^{-1}(\dd t + \omega),  &  &
	A^{(2)} =\dd t  - h_2^{-1}(\dd t + \omega).
\end{align}
It is useful to write the vector dual to the B-field. We define\footnote{The minus sign in \eqref{A3-def} ensures that the sign of the Chern-Simons term of the resulting U(1)$^3$ supergravity is same as what is used in the Bena-Warner literature.}, 
\be \label{A3-def}
e^{-\frac{\sqrt{2}}{\sqrt{3}}\chi - 2\Phi} \star_5 H =: - d A^{(3)}. 
\ee
We find, 
\be
A^{(3)} = \frac{1}{2} q (1 - x) \dd \phi. 
\ee

The angular momentum in the BPS limit is related to the dipole charge via,
\begin{equation}
J := J_\psi=\frac{\pi }{4 G_5} (2 \varkappa^2) q.
\end{equation}
Since $q$ has both an upper and a lower bound,  $J$ also has both an upper and a lower bound.

The horizon in the BPS limit recedes to $y \to -\infty$. The area of the horizon is zero, hence the name small black ring.  The angular velocity of the horizon is also zero, as is expected for a supersymmetric black hole. 

This BPS solution can  readily be written in the Bena-Warner \cite{Bena:2007kg} form (see appendix \ref{sec:appendix-BW} for notation). To do so, we first write the four-dimensional flat base space \eqref{ds4} in a standard set of coordinates. Define,
\begin{align}
    r_1 &= \sqrt{2} \varkappa  \frac{\sqrt{1-x^2}}{x-y}, &
    r_2 &= \sqrt{2} \varkappa  \frac{\sqrt{y^2-1}}{x-y}.
\end{align}
In these coordinates, flat space metric \eqref{ds4} becomes 
\be
\dd s^2_4 = \dd r_1^2 + r_1^2 \dd \phi^2 + \dd r_2^2 + r_2^2 \dd \psi^2.  
\ee
Next we note that the function $\Sigma^{-1}$,
\be
\Sigma^{-1} = \frac{1}{4\varkappa^2}(x-y), 
\ee
solves the Laplace equation on four-dimensional flat space for a ring source at $r_1 =0, r_2= \sqrt{2} \varkappa , 0 \le \psi <2\pi$ \cite{Emparan:2006mm}. We now do the following standard series of coordinate transformations. First,
\begin{align}
r_1 &= \rho \cos \Theta, &
r_2 &= \rho \sin \Theta,  &
\phi &= \frac{1}{2} \left(\phi_1  + \phi_2\right)&
\psi &= \frac{1}{2} \left(\phi_1  - \phi_2\right),
\end{align}
then,
\begin{align}
\Theta &= \frac{1}{2} \theta, &
\rho &= 2 \sqrt{r},
\end{align}
and finally,
\begin{align}
x_1 &= r \sin \theta \cos \phi_2, &
x_2 &= r \sin \theta \sin \phi_2, &
x_3 &= r \cos \theta.
\end{align}
The four-dimensional flat base space is now written as
\be
\dd s^2_4 = r ( \dd \phi_1 + \cos \theta \dd \phi_2)^2 + \frac{1}{r} ( \dd r^2  + r^2 \dd \theta^2 + r^2 \sin^2 \theta \dd \phi_2^2), \label{4d-flat-GH}
\ee
In these coordinates, 
\be
\frac{1}{4}\Sigma =  r_o := | \vec x - \vec x_o|,  \qquad  \mbox{where} \qquad  \vec x_o = \left(0,0, -\frac{1}{2} \varkappa^2\right).
\ee
From this discussion it is clear that the Bena-Warner\footnote{A concise review of the Bena-Warner formalism can be found in \cite{Adhikari:2024zif}. For the three-dimensional Hodge dualities, we use conventions such that for  $V=1/r$, $ \dd V = \star_3 \dd A$ implies $A = \cos \theta \dd \phi_2$, i.e., $\epsilon_{r \theta \phi_2} > 0$. For actual calculations, we need to translate this in $x,y$ coordinates.} harmonic function $V$ is simply $1/|\vec x|$. With a little bit of work, we can figure out the remaining harmonic functions. 
We find, 
\begin{align}
L_1 &= 1 + \frac{Q_1}{4 r_o}, & K^1 &=0,  &
L_2 &= 1 +  \frac{Q_2}{4 r_o},&  K^2 &=0, \\
L_3 &= 1 & K^3 &= \frac{q}{2 r_o}, &
V &= \frac{1}{r}, & M &=- \frac{q}{4}+  \frac{q \varkappa^2}{8r_o}.
\end{align}
We note that $h_{1,2} = L_{1,2}$. Putting $\alpha = 1$ in eq.~\eqref{dipole-alpha}\footnote{Recall that $\alpha =1$ corresponds to the situation when the seed solution has no dipole charge,  cf.~\eqref{introducing-alpha}.}, the dipole charge takes its minimum allowed value, 
\be
q = \frac{1}{2\varkappa}\sqrt{Q_1 Q_2}.
\ee
In this limit, the eight harmonic functions perfectly match with equations (6.21) and (6.22) of \cite{Bandyopadhyay:2025jbc}. This is a  non-trivial consistency check.

\subsection{Near horizon geometry of the small black ring}

In this section, we revisit the near horizon geometry of the small black ring used in the scaling analysis of \cite{Dabholkar:2006za}. We obtain the near horizon geometry from the Bena-Warner harmonic functions. The analysis is almost the same as there, though the emphasis is somewhat different. This change in emphasis is relevant for our upcoming work \cite{Dharanipragada:2026dji},  where we discuss a similar limit for the BPS version of the solution presented in section \ref{sec:Teo}.

It is most convenient to work in the six-dimensional string frame \eqref{action_6sd}. In our presentation so far, the asymptotic value of the dilaton $\Phi$ has been set to zero. For the scaling analysis, the string coupling $g$ needs to be restored, so that the asymptotic dilaton goes as $e^\Phi \to g$. Since shifting $\Phi$ by a constant is a symmetry of the string frame equations of motion, this can be simply achieved by multiplying $e^\Phi$ by a factor of $g$. The 
 Einstein frame metric
can then be related to the string frame metric via,
\be\label{edefcan}
G^{(E)}_{MN} =g e^{-\Phi} G^{(S)}_{MN}.
\ee
 The  factor of $g$  was not included in \eqref{eq:string-Einstein}, but can be included to ensure $G^{(E)}_{MN}$  approaches $\eta_{MN}$ asymptotically, as does $G^{(S)}_{MN}$. In this section, we shall always work with the six-dimensional string frame metric. For the configurations discussed in this paper,  
the six-dimensional string frame metric takes the form,
\be
\dd s^2_{6S} =  e^{\frac{1}{\sqrt{6}}\chi + \Phi} \dd s_{5E}^2 + e^{-\frac{\sqrt{3}}{\sqrt{2}} \chi + \Phi} (\dd z + A^{(1)})^2. 
\ee
Let the radius of the $z$ circle be $R_z$. For the F1-P small black ring this metric is,
\be
\dd s^2_{6S} = -\frac{1}{h_1 h_2} ( \dd t + \omega)^2  + \frac{h_1}{h_2} \left(\dd z + \dd t - h_1^{-1}(\dd t + \omega) \right)^2+ \dd s^2_4. 
\ee
We can rewrite the metric in a more convenient form as,
\be \label{6d-string-final}
\dd s^2_{6S} = \frac{1}{h_2} \left\{-( \dd t + \omega)^2  + (\dd z -\omega)^2 + (h_1 -1) (\dd t + \dd z)^2 \right\} + \dd s^2_4. 
\ee

In the Bena-Warner description, we saw that the black ring is located at $\vec x = \vec x_o$.  To zoom in near the stretched horizon of the ring, we  take 
\be
r_o \ll Q_1, Q_2, \varkappa^2. \label{inequalities}
\ee
In this limit, the harmonic functions behave as,
\begin{align} \label{limit-1}
L_1 & \simeq \frac{Q_1}{4 r_o}, & K^1 &=0,  &
L_2 &\simeq  \frac{Q_2}{4 r_o},&  K^2 &=0, \\
L_3 & = 1 & K^3 &= \frac{q}{2 r_o}, &
V &\simeq \frac{2}{\varkappa^2}, & M &\simeq \frac{q \varkappa^2}{8r_o}.\label{limit-2}
\end{align}

Since the Bena-Warner function $V$ has become a constant in this limit, the four dimensional base space becomes $\mathbb{R}^3 \times S^1$. The base metric takes the form, 
\be
\dd s^2_4 \simeq \frac{\varkappa^2}{2} \dd \psi_o^2 + \frac{2}{\varkappa^2} \left(d r_o^2 + r_o^2 \dd \theta_o^2 + r_o^2  \sin^2 \theta_o \dd \phi_o^2\right),
\ee
where $(r_o, \theta_o, \phi_o)$ are the spherical polar coordinates centered at $\vec x = \vec x_o$.  These coordinates should not be confused with $(r, \theta, \phi_2)$ used above centered at $\vec x = 0$. In the ring coordinates  \eqref{ds4}, the $\frac{r_o}{\varkappa^2} \ll 1 $ limit corresponds to $- y \gg 1$, keeping other coordinates fixed.
From there, we can identify that $x = \cos \theta_o$ and $\phi = \phi_o$ and $\psi = \frac{1}{2} \psi_o$. Now, we define 
\be
\rho = \frac{\sqrt{2}}{\varkappa} r_o,
\ee
so that 
\be \label{base-simp}
\dd s^2_4 \simeq 2\varkappa^2 \dd \psi^2 + \dd \rho^2 + \rho^2 ( \dd \theta_o^2 + \sin^2 \theta_o \dd \phi^2).
\ee
We will shortly see that this $\rho$ is the same radial coordinate that features in the analysis of \cite{Dabholkar:2006za}. From \eqref{base-simp}, we also see that $\sqrt{2} \kappa$ sets the size of the ring. This unusual factor of $\sqrt{2}$ is now standard in the literature.

Through the Bena-Warner formalism we can readily compute the one-form $\omega$ in the limit where the harmonic functions take the form \eqref{limit-1}--\eqref{limit-2}. We have, 
\be
\omega = \left(\frac{K^3}{2V} + M\right) \dd \psi_o + \omega_3. 
\ee
To find $ \omega_3$ we need to use the duality relation \eqref{omega-2}. In the limit where the harmonic functions are \eqref{limit-1}--\eqref{limit-2}, we have
\bea
& & \star_3 \dd \omega_3 = V \dd M - \frac{1}{2} \dd K^3 = 0, \\ \implies & & \omega_3 = 0. 
\eea
Therefore, we simply have, 
\be \label{omega-simp}
\omega \simeq \frac{q \varkappa^2}{4 r_o}\dd \psi_o = \frac{q \varkappa^2}{2 r_o}\dd \psi. 
\ee
Inserting \eqref{base-simp} and \eqref{omega-simp} in \eqref{6d-string-final}, we get the metric near the stretched horizon of the ring. In this metric, the $S^2$ cross-section of the ring  is  simply $\rho^2 ( \dd \theta_o^2 + \sin^2 \theta_o \dd \phi^2)$. 
For the $S^1$ of the ring we have $2\varkappa^2 \dd \psi^2$.  $\alpha' = 1$ in our conventions. The horizon of the ring is located at $\rho = 0$. The curvature and the other field strengths associated with the near-horizon configuration are small only for $\rho \gg 1$.  Thus, for the higher derivative corrections to be negligible we require $\rho \gg 1$. Moreover, for the ring to be macroscopic we require $\varkappa^2 \gg 1$.

The parameters $Q_1, Q_2, q$ and $\varkappa^2$ are related to the quantized charges $n,w,Q$ and angular momentum $J$ via the relations \cite{Dabholkar:2006za} (with $\alpha' = 1$),
\begin{align} \label{quantized}
 q& = \frac{g^2 Q}{R_z}, & \varkappa^2 &= \frac{J}{2Q}, & Q_1 &= \frac{g^2 n}{R_z^2}, & Q_2 &= g^2 w.
\end{align}
Here $n$ and $w$ denote the integer units of momentum and winding charge along the $S^1$ labeled by $z$.  $Q$ represents the integer units of winding charge along the $S^1$ of the ring. The upper bound in the inequality \eqref{bound} translates into 
\be
n w - JQ \ge 0. 
\ee

We work with the following choice of charges,
\begin{align} \label{charge-choices}
& J \gg Q \gg 1, & & n \sim w, & & n w \sim J Q, & & 1- \frac{JQ}{nw} \sim 1.
\end{align}
Through \eqref{quantized}, the first  condition in \eqref{charge-choices} ensures that $\varkappa^2 \gg 1$, which is one of the requirements of the near horizon limit. The other requirements, cf.~\eqref{inequalities}, are met through \eqref{charge-choices} when 
\be
\rho \ll g^2 Q, \frac{g^2Q}{R_z^2}, \varkappa. 
\ee
We define, 
\begin{align}
\sigma &= \sqrt{\frac{n}{w} - \frac{JQ}{w^2}} \frac{1}{R_z} (z + t),\\
\tau &= \frac{2\sqrt{2} \varkappa R_z}{g^2 \sqrt{nw - JQ} } t, \\
\chi &= \sqrt{\frac{J}{Q}} \psi - \frac{\sqrt{JQ}}{w R_z} (z + t),
\end{align}
and the metric \eqref{6d-string-final} in the near horizon region becomes,
\be
\dd s^2 \simeq \dd \sigma^2 +  \dd \chi^2 - 2 \rho \dd \tau \dd \sigma + \dd \rho^2 + \rho^2 ( \dd \theta_o^2 + \sin^2 \theta_o \dd \phi^2).
\ee
The dilaton becomes, 
\be
e^{2 \Phi} = \frac{g^2}{h_2} \simeq 2 \sqrt{\frac{J}{Q}} \frac{\rho}{ w},
\ee
and the B-field (up to constant additive terms) becomes,
\be
B \simeq - \rho \, \dd \tau \wedge \dd \sigma.
\ee
The metric is singular in the $\rho \to 0$ limit. 

The coordinates $\sigma$ and $\chi$  have the following periodic identifications,
\be
(\sigma, \chi) \equiv \left( \sigma, \chi + 2 \pi \sqrt{\frac{J}{Q}}\right) \equiv \left(\sigma + 2\pi \sqrt{\frac{n}{w}}\sqrt{1 - \frac{JQ}{nw}}, \chi - 2\pi \frac{\sqrt{JQ}}{w} \right). 
\ee
The region where the above solution is valid is, 
\be
1 \ll \rho \ll g^2 Q, \frac{g^2Q}{R_z^2}, \varkappa.  
\ee
The above form of the solution is precisely what is given in \cite[eqs.~(3.21)--(3.23)]{Dabholkar:2006za}. The solution and the periodicities of the $(\sigma, \chi)$ coordinates are independent of the parameters $g$ and $R_z$.  

\section{Doubly spinning non-supersymmetric F1-P black ring}
In this section, we use the Chen-Hong-Teo \cite{Chen:2012kd} doubly spinning dipole black ring as the seed to generate a two-charge doubly-spinning dipole black ring, following the procedure discussed in section \ref{sec:dualities}.
\label{sec:Teo}
\subsection{Chen-Hong-Teo dipole black ring}
In coordinates $(t,y,x,\psi,\phi)$,  the metric for the Chen-Hong-Teo solution  takes the form
\begin{align}
	\dd s_5^2=&-\bigg[\frac{H(y,x)^3}{K(x,y)^2H(x,y)}\bigg]^{1/3}(\dd t+\omega_\psi \dd \psi +\omega_\phi \dd \phi)^2 + \frac{2\varkappa^2}{(x-y)^2}[K(x,y)H(x,y)^2]^{1/3} \nonumber\\[2pt]
	&\times\biggl\{\frac{F(x,y)}{H(x,y)H(y,x)}\, (\dd \psi +\omega_{\psi \phi}\dd \phi)^2-\frac{G(x)G(y)}{F(x,y)}\,\dd\phi^2+\frac{1}{UV}\biggl[\frac{\dd x^2}{G(x)}-\frac{\dd y^2}{G(y)}\biggr]\biggr\}.
\end{align}
The functions $\omega_{\psi},\omega_\phi,\omega_{\psi\phi}$ are
\begin{align}
	\omega_{\psi}&=\sqrt{\frac{2a(a+c)}{UV}}\frac{\varkappa(1+b)(1+y)J_+(x,y)}{H(y,x)},\\[2pt]
	\omega_\phi &=\sqrt{\frac{2ab(a+c)(1-a^2)}{UV}}\frac{\varkappa c(1+b)(1-x^2)}{H(y,x)}[(1+cy)(a+ab+by)-c-y],\\[2pt]
	\omega_{\psi\phi}&=\frac{\sqrt{b(1-a^2)}}{UV}\frac{ac(1+b)(x-y)(1-x^2)(1-y^2)}{F(x,y)}\nonumber\\[2pt]
	&\times[b(1+cx)(1+cy)(1-b-a^2-a^2b)-(1-c^2)(1-b+a^2+a^2b)],
\end{align}
and the functions $G,K,H,F$ and $J_+$ are,
\begin{align}\label{eq:Teofunc}
&	G(x)= (1-x^2)(1+cx)\,, \\[3pt]
&	K(x,y)= -a^2(1+b)\left[bx^2(1+cy)^2+(c+x)^2\right]+\left[b(1+cy)-1-cx\right]^2+bc^2(1-xy)^2,
\end{align}
\begin{align}\label{eq:Teofunc2}
	H(x,y)=&-a^2(1+b)\left[b(1+cx)(1+cy)xy+(c+x)(c+y)\right]-a(1+b)(x-y)\big[c^2-1 \nonumber \\[2pt]
	&+b(1+cx)(1+cy)\big]+\left[b(1+cy)-1-cx\right]\left[b(1+cx)-1-cy\right]+bc^2(1-xy)^2,  \\[2pt]
	F(x,y)=&\frac{1-y^2}{UV}\,\bigg\{bcG(x)\Big\{c(y^2-1)\left[a^2(1+b)-b+1\right]^2-4a^2y(1-b^2)(1+cy)\Big\} \nonumber \\[2pt]
	&-(1+cy)\Big\{a^2(1+b)^2\left[a^2(c+x+bx+bcx^2)^2-(c+x-bx-bcx^2)^2\right] \nonumber \\[2pt]
	&-(1-b)^2(1+cx)^2\left[a^2(1+b)^2-(1-b)^2\right]\Big\}\bigg\}\,
\end{align}
\begin{align}\label{eq:Teofunc3}
	J_\pm(x,y)=&a^2(1+b)\left[bx(1+cx)(1+cy)+(1+c)(c+x)\right] \nonumber \\[2pt]
	&\pm a\left\{(1-x)\left[b(1+cx)+c-1\right]\left[b(1+cy)+c+1\right]-2bc(1-y)(1+cx)\right\} \nonumber \\[2pt]
	&-\left[b(1+cx)-c-1\right]\left[b(1+cy)-cx-1\right]-bc^2(1-x)(1-xy)\,.
\end{align}
The $J_-$ function will be used shortly. The angular coordinates $(\psi, \phi)$ are canonically normalised. The radial coordinate $y$ takes the range $- \infty < y \le -1$, and the coordinate $x$ takes the range $- 1 \le x \le 1$. 

There are four free parameters in the solution 
\be
a,b,c, \varkappa.
\ee
$U,V$ are given in terms of parameters $a, b$ as,
\begin{align}
	U&=1+a-b+ab,& V&=1-a-b-ab.
\end{align}
A physical interpretation of these parameters is as follows: $\varkappa$ sets the scale of the solution, $a-c$ is related to the dipole charge, $b$ is related to the rotation on the $S^2$, and $c$ is related to the size of the black hole horizon. The horizon is at $y=-1/c$.  For the analysis that follows, two observations are particularly important:
\begin{enumerate}
\item Setting $b=0$ gets rid of the rotation on the $S^2$. We get back Emparan's singly spinning  dipole black ring solution discussed at the beginning of  section \ref{sec:EEF}. 
\item Setting $a=c$ gets rid of the dipole charge. We get the Pomeransky-Sen'kov doubly spinning black ring \cite{Pomeransky:2006bd}. 
\end{enumerate}
We will have more to say about  these limits in section \ref{sec:limits}.

The four independent parameters are subjected to the constraints 
\begin{align}
& 0 \le c \le a < 1,  & & 0 \le b < \frac{1-a}{1+a}, & & \varkappa > 0.
\end{align}
These constraints ensure that the quantities $U$ and $V$ are positive. 
 
The B-field supporting the solution is\footnote{In our presentation, the over-all sign of the B-field is flipped compared to \cite{Chen:2012kd}. As far as the doubly spinning dipole black ring is concerned this sign is a convention.}
\begin{equation}
	B=B_{t\psi} \, \dd t\wedge\dd \psi +B_{t\phi} \, \dd t\wedge\dd \phi +B_{\phi\psi} \, \dd \phi\wedge\dd \psi ,
\end{equation}
with
\begin{align}
	B_{t\psi}=&\sqrt{\frac{2a(a-c)}{UV}}\frac{\varkappa (1+b)(1+y)J_{-}(x,y)}{H(x,y)}\,,  \\[3pt]
	B_{t\phi}=&\sqrt\frac{2ab(1-a^2)(a-c)}{UV}\frac{c\varkappa(1+b)(1-x^2)[c+y+(1+cy)(a+ab-by)]}{H(x,y)}\,,  \\[3pt]
	B_{\phi\psi}=&-\frac{2c\varkappa^2(1+b)\sqrt{b(1-a^2)(a^2-c^2)}}{V\times(x-y)H(x,y)}\,(1-x^2)(1+y)[a(1+b)(1-x)(1+cy) \nonumber \\[3pt]
	&+(1-y)(b+bcy-1-cx)]\,.
\end{align}
The dilaton $\widetilde{\phi}$ is
\begin{equation}
	e^{\widetilde \phi} = e^{\frac{2\sqrt{2}}{\sqrt{3}} \Phi} =\left[\frac{K(x,y)}{H(x,y)}\right]^\frac{\sqrt{2}}{\sqrt{3}}.
\end{equation}

This cumbersome solution was constructed through a clever application of the inverse scattering method to vacuum six-dimensional gravity, as proposed in \cite{Rocha:2011vv}. 

\subsection{The charged solution}

\label{sec:2-charge-general}
Using the dualities  described in \ref{sec:dualities}, we can straightforwardly add F1 and P charges to the  Chen-Hong-Teo dipole black ring. The final five-dimensional Einstein frame metric can be written as,
\bea
  \mathrm{d}s^2_5 &= & - (h_1 h_2)^{-\frac{2}{3}}\left[ \frac{H(y,x)^3}{K(x,y)^2 H(x,y)} \right]^{\frac{1}{3}} (\mathrm{d} t +\widetilde{\omega}_\psi \mathrm{d} \psi 
   + \widetilde{\omega}_\phi \mathrm{d} \phi)^2 \nonumber \\[2mm]  && + (h_1 h_2)^{\frac{1}{3}}  \left[ K(x,y) H(x,y)^2 \right]^{\frac{1}{3}} \label{metric-final} \\[2mm]
 & & \times \frac{2 \varkappa^2}{(x-y)^2} \left\{ \frac{F(x,y) \left( \mathrm{d}\psi + \omega_{\psi\phi} \, \mathrm{d}\phi \right)^2}{H(x,y) H(y,x)} 
  - \frac{G(x) G(y) \, \mathrm{d}\phi^2}{F(x,y)}+ \frac{1}{U V} \left[ \frac{\mathrm{d}x^2}{G(x)} - \frac{\mathrm{d}y^2}{G(y)} \right] \right\}, \nonumber
\eea
where
\begin{equation}
  h_i=c_i^2-\frac{H(y,x)}{H(x,y)} s_i^2,  \qquad \mbox{for} \qquad i=1,2,
\end{equation}
and
\bea
  \widetilde{\omega}_\psi &=& c_1 c_2 \omega_\psi + s_1 s_2 B_{t\psi}, \\
  \widetilde{\omega}_\phi &=& c_1 c_2 \omega_\phi + s_1 s_2 B_{t\phi}.
\eea
The two-form supporting the solution is 
\begin{align}
  B = \widetilde{B}_{t\phi} \, \mathrm{d}t \wedge \mathrm{d}\phi + \widetilde{B}_{t\psi} \, \mathrm{d}t \wedge \mathrm{d}\psi + \widetilde{B}_{\phi\psi} \, \mathrm{d}\phi \wedge \mathrm{d}\psi,
\end{align}
where
\begin{align} \label{B-final}
  \widetilde{B}_{t\phi} & = \frac{1}{h_2} \left( c_1 c_2 B_{t\phi} + s_1 s_2 \omega_\phi \frac{H(y,x)}{H(x,y)}\right) ,  \\[2pt]
   \widetilde{B}_{t\psi} &= \frac{1}{h_2} \left( c_1 c_2 B_{t\psi}+ s_1 s_2 \omega_\psi \frac{H(y,x)}{H(x,y)} \right), \\[2pt]
  \widetilde{B}_{\phi\psi} & = \frac{1}{h_2}\left( c_2^2 B_{\phi\psi} - s_2^2(B_{\phi\psi}-B_{t\psi} \omega_{\phi} + B_{t\phi} \omega_{\psi}) \frac{H(y,x)}{H(x,y)}\right).
\end{align}
The two vector fields take the form
\begin{align} \label{vector-final-t}
  A_t^{(1)} & = h_1^{-1} \left(1-\frac{H(y,x)}{H(x,y)}\right)c_1 s_1, \\
  A_t^{(2)} & =  h_2^{-1} \left(1-\frac{H(y,x)}{H(x,y)}\right)c_2 s_2,  \\ 
  \label{vector-final-phi}
  A_{\phi}^{(1)} & =  -h_1^{-1}  \left(c_1 s_2 B_{t\phi} + s_1 c_2  \omega_\phi \frac{H(y,x)}{H(x,y)}\right), \\
    A_{\phi}^{(2)} & = -h_2^{-1} \left(c_2 s_1 B_{t\phi} +  s_2 c_1 \omega_\phi \frac{H(y,x)}{H(x,y)}\right), \\
      \label{vector-final-psi}
  A_{\psi}^{(1)} & = -h_1^{-1}  \left(c_1 s_2 B_{t\psi} + s_1 c_2  \omega_\psi \frac{H(y,x)}{H(x,y)}\right),  \\
    A_{\psi}^{(2)} & =- h_2^{-1} \left(c_2 s_1 B_{t\psi} +  s_2 c_1 \omega_\psi \frac{H(y,x)}{H(x,y)}\right). 
\end{align}
Finally, the scalars are
\begin{align} 
  e^{2\Phi} & =  h_2^{-1} \frac{K(x,y)}{H(x,y)} ,\\ 
  e^{- \sqrt{\frac{3}{2}} \chi} &=  \frac{h_1}{\sqrt{h_2}} \sqrt{\frac{H(x,y)}{K(x,y)}}. \label{scalars-final}
\end{align}

\subsection{Physical properties of the solution}
\label{sec:physical-properties-general}
The asymptotically flat nature of the solution  \eqref{metric-final}--\eqref{scalars-final} can be made manifest by changing coordinates as,
\bea
x&=& -1 + \frac{4 \varkappa^2}{r^2} (1-c) \cos^2 \theta, \\
y &=& -1 - \frac{4 \varkappa^2}{r^2} (1-c) \sin^2 \theta.
\eea
The ADM mass, angular momenta, and electric charges can then be readily calculated.  The ADM mass of the black ring is, 
\be
		M = \frac{\pi \varkappa^2}{G_5 U V} (1 + b) \left\{(a + c) U + a(1 - b + c + b c)  (\cosh 2 \delta_1 + \cosh 2\delta_2-1)\right\} .
        \ee
The $S^1$ angular momentum $J_\psi$ is, 
\begin{align}
    J_\psi &=  \frac{2\pi \varkappa^3 (1 + b)}{G_5}\sqrt{\frac{a}{2UV}}  \\
  &  \quad    \times   \left[\frac{\sqrt{a+c}}{V} \, \left\{2 (1+a) c + (1-c) U \right\}\, c_1 c_2   + \frac{\sqrt{a-c}}{U} \, \left\{ 2 (1-a)  c+(1-c) V \right\} \, s_1 s_2 \right],  \nonumber
\end{align}
and the $S^2$ angular momentum $J_\phi$ is, 
\be
J_\phi = \frac{2\pi \varkappa^3  (1 + b)c}{G_5}\sqrt{\frac{2ab \left(1-a^2\right) }{U V}} \left[\frac{\sqrt{a+c}}{V}c_1c_2- \frac{\sqrt{a-c}}{U}s_1s_2 \right].
\ee
The P- and F1- charges in the normalisation as in \eqref{charge-1}--\eqref{charge-2} are respectively
\bea
		\mathbf{Q}_1 & =& \frac{\pi  \varkappa^2}{G_5  U V} (2 a)  (1+b) \{ 1+ c - b(1-c)\} c_1 s_1, \\
		\mathbf{Q}_2 & =& \frac{\pi  \varkappa^2}{G_5  U V} (2 a)  (1+b) \{ 1+ c - b(1-c)\} c_2 s_2.
\eea
The dipole charge in the normalisation \eqref{dipole-charge-EEF} is 
\be 
q = \frac{2\varkappa(1+b)}{\sqrt{U\,V}} \left[\sqrt{2 a (a-c)} \, c_1 c_2 + \sqrt{2 a (a+c)} \, s_1 s_2 \right].
\ee
The horizon is at $y=-1/c$.  The horizon area, horizon temperature, horizon angular velocities\footnote{We found it easiest to compute these quantities following appendix A of \cite{Astefanesei:2010bm}.} are, respectively,
\bea 
&& {\cal A}_H = 16\pi^2 \varkappa^3 c (1 + b) \sqrt{\frac{2 a \left(1-a^2\right)}{U V}} \left[\frac{\sqrt{a+c}}{V} c_1 c_2 - \frac{\sqrt{a-c}}{U} s_1 s_2\right],\\[2mm]
& &T = \sqrt{\frac{U V}{2 a \left(1-a^2\right)}} \left[ \frac{U V}{4 \pi  (1+b)  \varkappa \left(c_1 c_2 U \sqrt{a+c} - s_1 s_2 V \sqrt{a-c} \right)} \right],\\[2mm]
& & \Omega_\psi  = \frac{\sqrt{a U V}}{\sqrt{2} \varkappa \left(c_1 c_2 U \sqrt{a+c} - s_1 s_2 V \sqrt{a-c} \right)}, \\[2mm]
& &\Omega_\phi  = \sqrt{\frac{b U V}{2 a \left(1-a^2\right)}} \left[ \frac{a (1+ a) (1+ b)+V}{(1+b) \varkappa \left(c_1 c_2 U \sqrt{a+c} - s_1 s_2 V \sqrt{a-c} \right)} \right].
\eea

\subsection{Limits}
\label{sec:limits}
\subsubsection{Recovering the singly spinning solution}
The $S^2$ rotation of the Chen-Hong-Teo solution is switched off when the parameter $b$ is set to zero.  Accordingly,  setting $b=0$ in  the solution of section \ref{sec:2-charge-general} reproduces the singly spinning solution as presented in section \ref{sec:EEF}. In this limit, the physical quantities  listed in section \ref{sec:physical-properties-general} reduce to those listed in section \ref{sec:EEF}.

\subsubsection{Recovering the charged solution without an independent  dipole charge}
\label{sec:limit-without-dipole}

The dipole charge of the Chen-Hong-Teo solution is switched off when the parameter $a$ is set equal to $c$. In this limit, the B-field vanishes and the dilaton becomes constant and decouples. The metric then reduces to the Pomeransky--Sen'kov black ring \cite{Pomeransky:2006bd}. The explicit form of the metric in the coordinates used above can be found in \cite{Chen:2012kd}. To obtain the solution in the form given in \cite{Pomeransky:2006bd}, one must perform certain parameter redefinitions and coordinate transformations, which are described in \cite{Chen:2012kd, Chen:2011jb}. Here we present transformations that allow us to relate to the charged solution of  \cite{Bandyopadhyay:2025jbc}.

The charged solution in~\cite{Bandyopadhyay:2025jbc} was constructed by applying dualities on the Pomeransky--Sen'kov black ring. The final solution carries a dipole charge, but \emph{not as an independent parameter.} This is because the seed solution used in that construction does not itself carry a dipole charge. The transformations that relate the charged solution of \cite{Bandyopadhyay:2025jbc} to the charged solution of the present paper, upon setting 
$a=c$ are,
\begin{align}
b&=\frac{\widetilde \nu(1-\widetilde \mu^2)}{\widetilde \mu(1-\widetilde \nu^2)}\,,& c&=\frac{\widetilde \mu-\widetilde \nu}{1-\widetilde \mu \widetilde \nu}\,, \label{para1} \\[2mm]
 x&=\frac{\widetilde{x}+\widetilde \nu}{1+ \widetilde \nu \widetilde {x}}\,,& y&=\frac{\widetilde{y}+\widetilde \nu}{1+\widetilde \nu \widetilde{y}}\,,
\end{align}
where the coordinates $\widetilde{x}$ and $\widetilde{y}$ are identified with the coordinates $x$ and $y$ of \cite{Bandyopadhyay:2025jbc}, respectively, and the parameters  $\widetilde \mu$ and $\widetilde \nu$ are related to parameters $\nu$ and $\eta$ used there via,
\bea
\nu &=& \widetilde \mu + \widetilde \nu, \label{para2} \\
\eta &=&  \widetilde \mu  \widetilde \nu. \label{para3}
\eea
The parameter $k$ there is the same as $\varkappa$ here. The boost parameters $\delta_1$ and $\delta_2$ are also the same. The parameter relations \eqref{para1}–\eqref{para3} play a crucial role in understanding the BPS limit of the doubly spinning charged solution relevant for the construction of the index saddle, which will be discussed in our forthcoming work \cite{Dharanipragada:2026dji}.

\subsection{\texorpdfstring{Extremal limit with $S = 2 \pi J_\phi$}{Extremal limit with S = 2 pi J phi}}

The solution of section \ref{sec:2-charge-general} admits a variety of extremal limits. Broadly speaking, it can reach extremality in three distinct ways (and combinations thereof): (i) by maximizing its conserved electric charges while holding other parameters fixed, (ii) by maximizing its dipole charge, and (iii) by maximizing its 
$S^2$ angular momentum. An exhaustive analysis of all possible  extremal limits is not something we are interested in.  Moreover, we are not interested here in the limit associated with maximizing the conserved electric charges, which we will instead consider in \cite{Dharanipragada:2026dji}, as it is closely tied to the supersymmetric limit. In the present paper, we focus on extremal limits of type (ii) and (iii), as well as combinations thereof. These limits are closely related to the extremal limit discussed in \cite{Chen:2012kd}.

We define,
\bea
\alpha &=& \frac{c}{2a}, \\
\beta &=& \frac{c}{1-b},
\eea
and take $c, a \to 0$ and $b \to 1$ while keeping $\alpha, \beta$ fixed. The resulting parameters satisfy $0 < \beta < \alpha \le 1/2$. The full extremal solution can be readily obtained from the various expressions given above. 

In this limit, the horizon is located at 
$
y = - \infty.
$  The horizon is regular and has finite area. The horizon temperature vanishes in this limit. The entropy and the $J_\phi$ angular momentum then become,
\bea
S &=& \frac{A}{4G_5} = \frac{4 \pi^2\varkappa^3 \alpha \beta^2}{G_5}\left( c_1 c_2 \sqrt{\frac{2(1+ 2 \alpha)}{(\alpha -\beta )^3 (\alpha +\beta )}}- s_1 s_2 \sqrt{\frac{2(1-2 \alpha)}{(\alpha -\beta ) (\alpha +\beta )^3}} \right), \\
J_\phi &=&  \frac{2 \pi\varkappa^3 \alpha \beta^2}{G_5}\left( c_1 c_2 \sqrt{\frac{2(1+ 2 \alpha)}{(\alpha -\beta )^3 (\alpha +\beta )}}- s_1 s_2 \sqrt{\frac{2(1-2 \alpha)}{(\alpha -\beta ) (\alpha +\beta )^3}} \right),
\eea
with the entropy the angular momentum satisfying 
\be
S = 2 \pi J_\phi. \label{S-J}
\ee
This relation holds for $\alpha = 1/2$ too, i.e., when the seed solution has zero dipole charge. Although this observation was not mentioned in \cite{Bandyopadhyay:2025jbc}, it can be readily verified from the expressions given there. The relation \eqref{S-J} was first observed for the extremal Pomeransky–Sen’kov black ring in ref.~\cite{Reall:2007jv}. Generically, such black rings possess a non-trivial ergoregion.

The extremal limit that corresponds to maximizing  the dipole charge with no rotation present on the $S^2$ can be achieved in two different ways. One can first set $b=0$  and then take $c \to 0$ keeping $a$ fixed. Alternatively, one can take $\alpha \to 0$ while  keeping $\beta/\alpha$ fixed. In both cases, one recovers the extremal \emph{singly} spinning dipole black ring with two electric charges. This black ring does not possess a smooth horizon, i.e.,  $S = 2 \pi J_\phi =0$. The parameter $a$ in the first limit is the same as $\beta/\alpha$ in the second limit.

\section{Conclusions and future directions}
\label{sec:conclusions}

In this paper, we have presented a smooth, Lorentzian, non-extremal, two-charge, doubly spinning dipole black ring solution. Since the doubly spinning dipole black ring is itself a technically involved solution, constructing the corresponding charged configuration is a non-trivial task. We have analyzed several important properties of the resulting two-charge, doubly spinning dipole black ring; however, our study is by no means exhaustive. Many further directions remain open, including an analysis of the first law, the Smarr relation, the near-horizon limit of the extremal black ring, and the associated phase diagram. Such investigations would take us well beyond the scope of the present work and are therefore left for future study. Our primary motivation is instead to provide the necessary Lorentzian non-extremal solution required for the construction of the gravitational index saddle for the supersymmetric F1–P black ring.
In forthcoming work \cite{Dharanipragada:2026dji}, we analyze the analytic continuation that yields the index saddle for the supersymmetric F1–P black ring, closely following the approach of \cite{Bandyopadhyay:2025jbc}.

\bigskip
\noindent \textbf{Acknowledgments:} The work of A.V. was partly supported by SERB Core Research Grant CRG/2023/000545. G.S.P. would like to thank Imtak Jeon and Robert de Mello Koch for arranging the visit at Huzhou University during the early stages of this work. The work of G.S.P. was supported by the National Natural Science Foundation of China (NSFC) under Grant No.~12247103.

\appendix

\section{Bena-Warner formalism}
\label{sec:appendix-BW}
To set the notation, it is useful to quickly review the Bena-Warner formalism~\cite{Bena:2007kg}.   The Bena-Warner solutions are written in terms of 8 harmonic functions $\{V, K^I, L_I, M\}$ to the five-dimensional U(1)$^3$ supergravity theory with the Lagrangian
\be
 {\cal L}_5 =  
 R \star 1 - G_{IJ} dX^I \wedge
\star dX^J - G_{IJ} F^I \wedge \star F^J - \frac{1}{6} C_{IJK} F^I
\wedge F^J \wedge A^K, 
\label{lag_5}
\ee
where  $G_{IJ} = \frac{1}{2}(X^I)^{-2} \delta_{IJ}$,  and $C_{IJK}=1$ if $(IJK)$ is a permutation of $(123)$ and $C_{IJK}=0$
otherwise.  The Maxwell field strengths are $F^I = dA^I$. The metric takes the form,
\be
\dd s^2 = -f^2(\dd t +\omega )^2+f^{-1} \dd s^2_{\mathrm{4d-base}}, \label{5d-metric}
\ee
with the four-dimensional base metric  $\dd s^2_{\mathrm{4d-base}}$ written in the Gibbons-Hawking form as,
\begin{equation}
	\dd s^2_{\mathrm{4d-base}} = V^{-1}(\dd \widetilde z+A)^2+ V\dd s^2_{\mathrm{3d-base}},
	\label{GibbHawk}
\end{equation}
with  three-dimensional base  $\dd s^2_{\mathrm{3d-base}}$ being flat and 
with the 1-form $A$ satisfying 
$ 
	\star_3 \dd A=\dd V,  
$ 
where $\star_3$ is the Hodge star in three-dimensions. $\widetilde z $ is the Gibbons-Hawking fiber coordinate; it should not be confused with the sixth dimension $z$.  The one-form $\omega$ on the four-dimensional base space is, 
\be
\omega = \mu(\dd \widetilde z +A) +\omega_3. \label{omega-1}
\ee
The function $\mu$ is given as, 
\be
\mu= \frac{1}{6}C_{IJK} \frac{K^IK^JK^K}{V^2}+\frac{1}{2V}K^IL_I +M,
\ee
and the three-dimensional one-form $\omega_3$ satisfies,
\be
\star_3 \dd \omega=V\dd M-M\dd V+\frac12(K^I\dd L_I-L_I\dd K^I).  \label{omega-2}
\ee
The function $f$ in equation \eqref{5d-metric} takes the form $ f=(h_1 h_2 h_3)^{-1/3} $ 
where the three functions $h_I$ are specified  as,
\be
h_I = \frac{1}{2V} C_{IJK} K^J K^K+L_I.
\ee
The scalars are $X^I = (f h_I)^{-1}.$ Finally, the three vectors are, 
\be
 A^I =- \frac{1}{h_I} (\dd t+k) + \frac{K^I}{V}(\dd \widetilde z + A)+\xi^I + \dd t, 
\ee
with the three-dimensional one-forms $\xi^I$ satisfying $\star_3\dd\xi^I=-\dd K^I.$ 

\bibliography{dipole-black-ring-1}

@article{Reall:2007jv,
    author = "Reall, Harvey S.",
    title = "{Counting the microstates of a vacuum black ring}",
    eprint = "0712.3226",
    archivePrefix = "arXiv",
    primaryClass = "hep-th",
    doi = "10.1088/1126-6708/2008/05/013",
    journal = "JHEP",
    volume = "05",
    pages = "013",
    year = "2008"
}

@article{Astefanesei:2010bm,
    author = "Astefanesei, Dumitru and Rodriguez, Maria J. and Theisen, Stefan",
    title = "{Thermodynamic instability of doubly spinning black objects}",
    eprint = "1003.2421",
    archivePrefix = "arXiv",
    primaryClass = "hep-th",
    reportNumber = "AEI-2010-039",
    doi = "10.1007/JHEP08(2010)046",
    journal = "JHEP",
    volume = "08",
    pages = "046",
    year = "2010"
}

@article{Adhikari:2024zif,
    author = "Adhikari, Soumya and Dharanipragada, Pavan and Goswami, Kaberi and Virmani, Amitabh",
    title = "{Attractor saddle for 5D black hole index}",
    eprint = "2411.12413",
    archivePrefix = "arXiv",
    primaryClass = "hep-th",
    doi = "10.1007/JHEP03(2025)180",
    journal = "JHEP",
    volume = "03",
    pages = "180",
    year = "2025"
}

@article{Sen:1995in,
    author = "Sen, Ashoke",
    title = "{Extremal black holes and elementary string states}",
    eprint = "hep-th/9504147",
    archivePrefix = "arXiv",
    reportNumber = "TIFR-TH-95-19",
    doi = "10.1142/S0217732395002234",
    journal = "Mod. Phys. Lett. A",
    volume = "10",
    pages = "2081--2094",
    year = "1995"
}

@article{Callan:1995hn,
    author = "Callan, Curtis G. and Maldacena, Juan Martin and Peet, Amanda W.",
    title = "{Extremal black holes as fundamental strings}",
    eprint = "hep-th/9510134",
    archivePrefix = "arXiv",
    reportNumber = "PUPT-1565",
    doi = "10.1016/0550-3213(96)00315-X",
    journal = "Nucl. Phys. B",
    volume = "475",
    pages = "645--678",
    year = "1996"
}

@article{Dabholkar:1995nc,
    author = "Dabholkar, Atish and Gauntlett, Jerome P. and Harvey, Jeffrey A. and Waldram, Daniel",
    title = "{Strings as solitons and black holes as strings}",
    eprint = "hep-th/9511053",
    archivePrefix = "arXiv",
    reportNumber = "CALT-68-2028, EFI-95-67, UPR-0681-T",
    doi = "10.1016/0550-3213(96)00266-0",
    journal = "Nucl. Phys. B",
    volume = "474",
    pages = "85--121",
    year = "1996"
}

@article{Dabholkar:1990yf,
    author = "Dabholkar, Atish and Gibbons, Gary W. and Harvey, Jeffrey A. and Ruiz Ruiz, Fernando",
    title = "{Superstrings and Solitons}",
    reportNumber = "PUPT-1163, DAMTP-R-90-1, EFI-90-11",
    doi = "10.1016/0550-3213(90)90157-9",
    journal = "Nucl. Phys. B",
    volume = "340",
    pages = "33--55",
    year = "1990"
}

@article{Dabholkar:1989jt,
    author = "Dabholkar, Atish and Harvey, Jeffrey A.",
    title = "{Nonrenormalization of the Superstring Tension}",
    reportNumber = "PUPT-1122",
    doi = "10.1103/PhysRevLett.63.478",
    journal = "Phys. Rev. Lett.",
    volume = "63",
    pages = "478",
    year = "1989"
}

@article{Lunin:2001fv,
    author = "Lunin, Oleg and Mathur, Samir D.",
    title = "{Metric of the multiply wound rotating string}",
    eprint = "hep-th/0105136",
    archivePrefix = "arXiv",
    reportNumber = "OHSTPY-HEP-T-01-014",
    doi = "10.1016/S0550-3213(01)00321-2",
    journal = "Nucl. Phys. B",
    volume = "610",
    pages = "49--76",
    year = "2001"
}

@article{Dabholkar:2004yr,
    author = "Dabholkar, Atish",
    title = "{Exact counting of black hole microstates}",
    eprint = "hep-th/0409148",
    archivePrefix = "arXiv",
    reportNumber = "SLAC-PUB-10717, SU-ITP-04-36, TIFR-TH-04-23",
    doi = "10.1103/PhysRevLett.94.241301",
    journal = "Phys. Rev. Lett.",
    volume = "94",
    pages = "241301",
    year = "2005"
}

@article{Emparan:2001wn,
    author = "Emparan, Roberto and Reall, Harvey S.",
    title = "{A Rotating black ring solution in five-dimensions}",
    eprint = "hep-th/0110260",
    archivePrefix = "arXiv",
    reportNumber = "CERN-TH-2001-294",
    doi = "10.1103/PhysRevLett.88.101101",
    journal = "Phys. Rev. Lett.",
    volume = "88",
    pages = "101101",
    year = "2002"
}

@article{Emparan:2004wy,
    author = "Emparan, Roberto",
    title = "{Rotating circular strings, and infinite nonuniqueness of black rings}",
    eprint = "hep-th/0402149",
    archivePrefix = "arXiv",
    doi = "10.1088/1126-6708/2004/03/064",
    journal = "JHEP",
    volume = "03",
    pages = "064",
    year = "2004"
}

@article{Elvang:2004rt,
    author = "Elvang, Henriette and Emparan, Roberto and Mateos, David and Reall, Harvey S.",
    title = "{A Supersymmetric black ring}",
    eprint = "hep-th/0407065",
    archivePrefix = "arXiv",
    reportNumber = "NSF-KITP-04-84",
    doi = "10.1103/PhysRevLett.93.211302",
    journal = "Phys. Rev. Lett.",
    volume = "93",
    pages = "211302",
    year = "2004"
}

@article{Emparan:2006mm,
    author = "Emparan, Roberto and Reall, Harvey S.",
    title = "{Black Rings}",
    eprint = "hep-th/0608012",
    archivePrefix = "arXiv",
    doi = "10.1088/0264-9381/23/20/R01",
    journal = "Class. Quant. Grav.",
    volume = "23",
    pages = "R169",
    year = "2006"
}

@article{Emparan:2008eg,
    author = "Emparan, Roberto and Reall, Harvey S.",
    title = "{Black Holes in Higher Dimensions}",
    eprint = "0801.3471",
    archivePrefix = "arXiv",
    primaryClass = "hep-th",
    doi = "10.12942/lrr-2008-6",
    journal = "Living Rev. Rel.",
    volume = "11",
    pages = "6",
    year = "2008"
}

@article{Elvang:2004xi,
    author = "Elvang, Henriette and Emparan, Roberto and Figueras, Pau",
    title = "{Non-supersymmetric black rings as thermally excited supertubes}",
    eprint = "hep-th/0412130",
    archivePrefix = "arXiv",
    doi = "10.1088/1126-6708/2005/02/031",
    journal = "JHEP",
    volume = "02",
    pages = "031",
    year = "2005"
}

@article{Dabholkar:2005qs,
    author = "Dabholkar, Atish and Iizuka, Norihiro and Iqubal, Ashik and Shigemori, Masaki",
    title = "{Precision microstate counting of small black rings}",
    eprint = "hep-th/0511120",
    archivePrefix = "arXiv",
    reportNumber = "TIFR-TH-05-44, CALT-68-2583",
    doi = "10.1103/PhysRevLett.96.071601",
    journal = "Phys. Rev. Lett.",
    volume = "96",
    pages = "071601",
    year = "2006"
}

@article{Dabholkar:2006za,
    author = "Dabholkar, Atish and Iizuka, Norihiro and Iqubal, Ashik and Sen, Ashoke and Shigemori, Masaki",
    title = "{Spinning strings as small black rings}",
    eprint = "hep-th/0611166",
    archivePrefix = "arXiv",
    reportNumber = "TIFR-TH-06-36, NSF-KITP-06-104, CALT-68-2617, HRI-P-06-11-001",
    doi = "10.1088/1126-6708/2007/04/017",
    journal = "JHEP",
    volume = "04",
    pages = "017",
    year = "2007"
}

@article{Sen:2009bm,
    author = "Sen, Ashoke",
    title = "{Two Charge System Revisited: Small Black Holes or Horizonless Solutions?}",
    eprint = "0908.3402",
    archivePrefix = "arXiv",
    primaryClass = "hep-th",
    doi = "10.1007/JHEP05(2010)097",
    journal = "JHEP",
    volume = "05",
    pages = "097",
    year = "2010"
}

@article{Elvang:2003mj,
    author = "Elvang, Henriette and Emparan, Roberto",
    title = "{Black rings, supertubes, and a stringy resolution of black hole nonuniqueness}",
    eprint = "hep-th/0310008",
    archivePrefix = "arXiv",
    doi = "10.1088/1126-6708/2003/11/035",
    journal = "JHEP",
    volume = "11",
    pages = "035",
    year = "2003"
}

@article{Cabo-Bizet:2018ehj,
    author = "Cabo-Bizet, Alejandro and Cassani, Davide and Martelli, Dario and Murthy, Sameer",
    title = "{Microscopic origin of the Bekenstein-Hawking entropy of supersymmetric AdS$_{5}$ black holes}",
    eprint = "1810.11442",
    archivePrefix = "arXiv",
    primaryClass = "hep-th",
    doi = "10.1007/JHEP10(2019)062",
    journal = "JHEP",
    volume = "10",
    pages = "062",
    year = "2019"
}

@article{Iliesiu:2021are,
    author = "Iliesiu, Luca V. and Kologlu, Murat and Turiaci, Gustavo J.",
    title = "{Supersymmetric indices factorize}",
    eprint = "2107.09062",
    archivePrefix = "arXiv",
    primaryClass = "hep-th",
    doi = "10.1007/JHEP05(2023)032",
    journal = "JHEP",
    volume = "05",
    pages = "032",
    year = "2023"
}

@article{Dharanipragada:2026dji,
    author = "Dharanipragada, Pavan and Punia, Gurmeet Singh and Virmani, Amitabh",
    title = "{Index saddle for supersymmetric F1-P black ring}",
    eprint = "2601.04624",
    archivePrefix = "arXiv",
    primaryClass = "hep-th",
    reportNumber = "USTC-ICTS/PCFT-26-04",
    month = "1",
    year = "2026"
}

@article{Bandyopadhyay:2025jbc,
    author = "Bandyopadhyay, Subhodip and Punia, Gurmeet Singh and Srivastava, Yogesh K. and Virmani, Amitabh",
    title = "{The gravitational index of a small black ring}",
    eprint = "2504.09982",
    archivePrefix = "arXiv",
    primaryClass = "hep-th",
    doi = "10.1007/JHEP07(2025)200",
    journal = "JHEP",
    volume = "07",
    pages = "200",
    year = "2025"
}

@article{Cassani:2025iix,
    author = "Cassani, Davide and Ruip{\'e}rez, Alejandro and Turetta, Enrico",
    title = "{Bubbling saddles of the gravitational index}",
    eprint = "2507.12650",
    archivePrefix = "arXiv",
    primaryClass = "hep-th",
    doi = "10.21468/SciPostPhys.19.5.134",
    journal = "SciPost Phys.",
    volume = "19",
    pages = "134",
    year = "2025"
}

@article{Boruch:2025sie,
    author = "Boruch, Jan and Emparan, Roberto and Iliesiu, Luca V. and Murthy, Sameer",
    title = "{Novel black saddles for 5d gravitational indices and the index enigma}",
    eprint = "2510.23699",
    archivePrefix = "arXiv",
    primaryClass = "hep-th",
    month = "10",
    year = "2025"
}

@article{Chowdhury:2024ngg,
    author = "Chowdhury, Chandramouli and Sen, Ashoke and Shanmugapriya, P. and Virmani, Amitabh",
    title = "{Supersymmetric index for small black holes}",
    eprint = "2401.13730",
    archivePrefix = "arXiv",
    primaryClass = "hep-th",
    doi = "10.1007/JHEP04(2024)136",
    journal = "JHEP",
    volume = "04",
    pages = "136",
    year = "2024"
}

@article{Chen:2024gmc,
    author = "Chen, Yiming and Murthy, Sameer and Turiaci, Gustavo J.",
    title = "{Gravitational index of the heterotic string}",
    eprint = "2402.03297",
    archivePrefix = "arXiv",
    primaryClass = "hep-th",
    doi = "10.1007/JHEP09(2024)041",
    journal = "JHEP",
    volume = "09",
    pages = "041",
    year = "2024"
}

@article{Gaiotto:2005xt,
    author = "Gaiotto, Davide and Strominger, Andrew and Yin, Xi",
    title = "{5D black rings and 4D black holes}",
    eprint = "hep-th/0504126",
    archivePrefix = "arXiv",
    doi = "10.1088/1126-6708/2006/02/023",
    journal = "JHEP",
    volume = "02",
    pages = "023",
    year = "2006"
}

@article{Gaiotto:2005gf,
    author = "Gaiotto, Davide and Strominger, Andrew and Yin, Xi",
    title = "{New connections between 4-D and 5-D black holes}",
    eprint = "hep-th/0503217",
    archivePrefix = "arXiv",
    doi = "10.1088/1126-6708/2006/02/024",
    journal = "JHEP",
    volume = "02",
    pages = "024",
    year = "2006"
}

@article{Gauntlett:2004wh,
    author = "Gauntlett, Jerome P. and Gutowski, Jan B.",
    title = "{Concentric black rings}",
    eprint = "hep-th/0408010",
    archivePrefix = "arXiv",
    doi = "10.1103/PhysRevD.71.025013",
    journal = "Phys. Rev. D",
    volume = "71",
    pages = "025013",
    year = "2005"
}

@article{Bena:2007kg,
    author = "Bena, Iosif and Warner, Nicholas P.",
    title = "{Black holes, black rings and their microstates}",
    eprint = "hep-th/0701216",
    archivePrefix = "arXiv",
    reportNumber = "SPHT-T07-008",
    doi = "10.1007/978-3-540-79523-0_1",
    journal = "Lect. Notes Phys.",
    volume = "755",
    pages = "1--92",
    year = "2008"
}

@article{Boruch:2025qdq,
    author = "Boruch, Jan and Emparan, Roberto and Iliesiu, Luca V. and Murthy, Sameer",
    title = "{The gravitational index of 5d black holes and black strings}",
    eprint = "2501.17909",
    archivePrefix = "arXiv",
    primaryClass = "hep-th",
    doi = "10.1007/JHEP06(2025)145",
    journal = "JHEP",
    volume = "06",
    pages = "145",
    year = "2025"
}

@article{Boruch:2025biv,
    author = "Boruch, Jan and Iliesiu, Luca V. and Murthy, Sameer and Turiaci, Gustavo J.",
    title = "{Multicentered black hole saddles for supersymmetric indices}",
    eprint = "2507.07166",
    archivePrefix = "arXiv",
    primaryClass = "hep-th",
    month = "7",
    year = "2025"
}

@article{Chen:2012kd,
    author = "Chen, Yu and Hong, Kenneth and Teo, Edward",
    title = "{A Doubly rotating black ring with dipole charge}",
    eprint = "1204.5785",
    archivePrefix = "arXiv",
    primaryClass = "hep-th",
    doi = "10.1007/JHEP06(2012)148",
    journal = "JHEP",
    volume = "06",
    pages = "148",
    year = "2012"
}

@article{Hassan:1991mq,
    author = "Hassan, S. F. and Sen, Ashoke",
    title = "{Twisting classical solutions in heterotic string theory}",
    eprint = "hep-th/9109038",
    archivePrefix = "arXiv",
    reportNumber = "TIFR-TH-91-40",
    doi = "10.1016/0550-3213(92)90336-A",
    journal = "Nucl. Phys. B",
    volume = "375",
    pages = "103--118",
    year = "1992"
}

@article{Maharana:1992my,
    author = "Maharana, Jnanadeva and Schwarz, John H.",
    title = "{Noncompact symmetries in string theory}",
    eprint = "hep-th/9207016",
    archivePrefix = "arXiv",
    reportNumber = "CALT-68-1790",
    doi = "10.1016/0550-3213(93)90387-5",
    journal = "Nucl. Phys. B",
    volume = "390",
    pages = "3--32",
    year = "1993"
}

@article{Pomeransky:2006bd,
    author = "Pomeransky, A. A. and Sen'kov, R. A.",
    title = "{Black ring with two angular momenta}",
    eprint = "hep-th/0612005",
    archivePrefix = "arXiv",
    month = "12",
    year = "2006"
}

@article{Rocha:2011vv,
    author = "Rocha, Jorge V. and Rodriguez, Maria J. and Virmani, Amitabh",
    title = "{Inverse Scattering Construction of a Dipole Black Ring}",
    eprint = "1108.3527",
    archivePrefix = "arXiv",
    primaryClass = "hep-th",
    reportNumber = "AEI-2011-056, ULB-TH-11-19",
    doi = "10.1007/JHEP11(2011)008",
    journal = "JHEP",
    volume = "11",
    pages = "008",
    year = "2011"
}

@article{Chen:2011jb,
    author = "Chen, Yu and Hong, Kenneth and Teo, Edward",
    title = "{Unbalanced Pomeransky-Sen'kov black ring}",
    eprint = "1108.1849",
    archivePrefix = "arXiv",
    primaryClass = "hep-th",
    doi = "10.1103/PhysRevD.84.084030",
    journal = "Phys. Rev. D",
    volume = "84",
    pages = "084030",
    year = "2011"
}
\bibliographystyle{JHEP}

\end{document}